\documentclass[12pt]{iopart}

\usepackage{graphicx}
\usepackage[usenames]{color}

\begin{document}


\title{Stable and mobile   
two-dimensional dipolar  ring-dark-in-bright  Bose-Einstein condensate
 soliton}

\author{ Adhikari S K \footnote{Adhikari@ift.unesp.br; URL: http://www.ift.unesp.br/users/Adhikari}
} 
\address{
Instituto de F\'{\i}sica Te\'orica, UNESP - Universidade Estadual Paulista, 01.140-070 S\~ao Paulo, S\~ao Paulo, Brazil
} 

\begin{abstract}

We demonstrate 
robust, stable, mobile    
two-dimensional (2D) dipolar ring-dark-in-bright (RDB) Bose-Einstein condensate (BEC) solitons     
   for repulsive  
contact interaction, subject to  a harmonic   trap along the $y$ direction perpendicular to the polarization direction $z$. Such a RDB soliton has a ring-shaped notch (zero in density)
imprinted on a 2D bright soliton
 free to move in the $x-z$ plane.
At medium  velocity the head-on collision of two such solitons is found to be quasi elastic with practically no deformation.  The possibility of creating the RDB soliton by phase imprinting is demonstrated.
  The findings are illustrated using numerical simulation  employing realistic   interaction parameters in a dipolar $^{164}$Dy BEC.

\end{abstract}


\maketitle

\section{Introduction}
 
A bright soliton is a localized peak in density  that can 
travel at a constant velocity in one dimension (1D) with its shape unchanged, due to a cancellation of nonlinear attraction and dispersive
effects \cite{rmp}.  Similarly, a dark soliton represents a dip or notch (zero) in density in a uniform medium
that can travel in 1D maintaining its shape. 
Bright solitons have been observed  in water waves, nonlinear optics, and Bose-Einstein condensate (BEC) etc. among others \cite{rmp}. 
In physical three-dimensional (3D) world, 1D bright solitons were   observed in BEC of $^7$Li \cite{1} and
$^{85}$Rb atoms \cite{3}.

In the absence of a uniform BEC in a laboratory to create a mobile 1D dark soliton, a notch  in density in a finite trapped quasi-one-dimensional
(quasi-1D) repulsive BEC
along its width, dividing the BEC into  two equal pieces, 
has been observed in a quasi-1D BEC of $^{87}$Rb atoms  \cite{dark,dark2,dark3}
and termed a dark soliton.
However, these dark solitons exhibit thermal and dynamical instabilities \cite{8,10}
except for very strong transverse traps corresponding to  a quasi-1D 
geometry.  For a moderate to weak transverse confinement, the dark solitons are 
unstable, show snake instability \cite{8} and decay to form a vortex ring \cite{dark3}.
Although the pioneering experiments on quasi-1D dark solitons \cite{dark} faced technical difficulties 
in their stabilization due to thermal and dynamical effects, these problems are somewhat under  control in recent investigations and     different aspects and properties of dark solitons are being studied lately \cite{expdark}. The dynamical instability of the dark solitons becomes explicit, when the quasi-1D condition is relaxed, and bending and snaking (dynamical) instability of the dark solitons seriously hamper both theoretical and experimental studies \cite{dark2,dark3,dlf}.

A more stable  variant of the dark soliton, called ring-dark soliton, has also  been suggested in a trapped repulsive BEC  \cite{rd} and in a uniform self-defocusing medium in
nonlinear optics \cite{rdo1} and later observed 
in optics \cite{rdo2}.  A  ring-dark soliton possesses
 a notch in density in the form of a ring   in a quasi-two-dimensional (quasi-2D) trapped  BEC thus creating a ring-shaped concentric  notch  in a disk-shaped BEC.  However,  the  ring-dark soliton created on a quasi-2D  BEC is    dynamically unstable in general \cite{rd} and this creates difficulty in their theoretical and experimental studies \cite{rdsins}. 
Generation of ring dark solitons by phase engineering and their oscillations in spin-1 Bose-Einstein condensates have also been studied theoretically \cite{spinor}.
This instability can be reduced by reducing the strength of the axial trap in a quasi-1D BEC and the in-plane trap in a quasi-2D BEC hosting the dark and ring-dark solitons, respectively.  
Nevertheless, the  ring-dark solitons are to be created in a repulsive quasi-2D  BEC and the in-plane trap  can not be arbitrarily reduced as this will lead 
to arbitrarily large BECs hosting these solitons, thus making it impossible to   observe them experimentally.

Recent 
observation of  dipolar BECs of $^{164}$Dy \cite{ExpDy,dy}, $^{168}$Er \cite{ExpEr} and 
  $^{52}$Cr \cite{cr} atoms 
 has opened new studies   in BEC solitons.
In a dipolar BEC, apart from the 1D solitons \cite{1D} free to move on a straight line,  a two-dimensional (2D) soliton \cite{2D} free to move  in a plane  is possible.  
Dipolar 1D and 2D BEC solitons can be realized either by a harmonic or an
optical-lattice trap
 in quasi-1D \cite{ol1D} and quasi-2D \cite{ol2D} settings, respectively.
Robust 1D and 2D  dipolar BEC solitons accommodating  a large number of atoms
are possible 
for a fully repulsive 
contact interaction \cite{1D}.   The long-range
dipolar attraction prevents the escape  of the atoms from the soliton and the 
contact repulsion gives stability against collapse.

We demonstrated that the dynamical instability of a  dark soliton can be eliminated if the dark soliton is imprinted on an 1D dipolar BEC soliton instead of on a trapped BEC \cite{dib1d}. 
Even a dynamically stable dark soliton can be realized on a 2D dipolar BEC soliton \cite{ska2}. These 
1D and 2D dipolar dark-in-bright solitons without any axial trap or an in-plane trap, respectively, 
do not exhibit bending or snake instability of nondipolar dark solitons in a repulsive trapped BEC. 

In this paper we demonstrate that a dynamically stable ring-dark soliton can be realized on a 2D  bright dipolar  soliton. Such a 2D dipolar ring-dark-in-bright (RDB) soliton,   harmonically trapped in the $y$ direction, is  
mobile in the $x-z$ plane with a constant velocity, where  $z$ is the polarization direction. These RDB solitons  are   stationary excitations
of the 2D dipolar bright solitons.

The statics and dynamics of the 2D dipolar RDB solitons are studied by a real-time propagation of the 2D and 3D
mean-field Gross-Pitaevskii (GP) equation.      The stability of these   solitons is established 
by studying their breathing oscillation over a long time subject to  a reasonable perturbation.  
At medium velocities,
the head-on collision between two RDB solitons in real-time simulation 
 is found to be quasi elastic for velocities along both $x$ and $z$ directions. 
   We also study 
the creation of the RDB solitons by phase imprinting \cite{dark,phase} a   2D Gaussian profile     by introducing  a phase difference of $\pi$ between  parts of  the BEC wave function    on both sides of the  dark ring.

\section{Mean-field Gross-Pitaevskii equations}

We consider a dipolar BEC soliton, with the 
mass, number of atoms, magnetic  dipole moment, and scattering length 
given by $m, N, 
\mu, a$, respectively.   The 
interaction between 
two atoms at  $\bf r$ and $\bf r'$ is  \cite{rpp}
\begin{eqnarray}\label{intrapot} 
V({\bf R})= 3
a_{\mathrm {dd}}V_{\mathrm {dd}}({\mathbf R})+4\pi a \delta({\bf R})
, \quad \bf R = (r-r'),
     \end{eqnarray}
 with \begin{eqnarray}  a_{\mathrm {dd}}=
\frac{\mu_0  \mu^2m}{12\pi \hbar ^2    }, \quad
V_{\mathrm {dd}}({\mathbf R})=\frac{1-3\cos^2 \theta}{{\bf R}^3},
\end{eqnarray}where 
  $\mu_0$ is the permeability of free space, 
$\theta$ is the angle made by the vector ${\bf R}$ with the polarization 
$z$ direction.  
This interaction  leads to {\it a fully asymmetric} stable,    2D  RDB soliton  
mobile in the $x-z$ plane
 in the presence of  a harmonic trap along $y$ axis.
The dimensionless GP equation in this case    
can be written as   \cite{rpp,dipolarCPC}
\begin{eqnarray}& \,
{ i} \frac{\partial \phi({\bf r},t)}{\partial t}=
{\Big [}  -\frac{\nabla^2}{2 }
+
\frac{1}{2} y^2
\nonumber \\  &  \,
+ g \vert \phi \vert^2
+ g_{\mathrm {dd}}
\int V_{\mathrm {dd}}({\mathbf R})\vert\phi({\mathbf r'},t)
\vert^2 d{\mathbf r}'  
{\Big ]}  \phi({\bf r},t),
\label{eq3}
\end{eqnarray}
where 
$g=4\pi a N,$ 
$g_{\mathrm {dd}}= 3N a_{\mathrm {dd}}$ and $\int |\phi({\bf r},t)|^2 d{\bf r}=1$. 
In  (\ref{eq3}), length is expressed in units of 
oscillator length  $l=\sqrt{\hbar/(m\omega)}$,  where $\omega$  is the circular frequency 
of the harmonic trap along  $y$ axis. The 
energy is in units of oscillator energy  $\hbar\omega$, probability density 
$|\phi|^2$ in units of $l^{-3}$, and time in units of $ 
t_0=1/\omega$. 
In    (\ref{eq3}) the dipolar interaction is anisotropic in the $x-z$ plane, 
thus making the 2D bright and RDB solitons anisotropic in this plane.  
 
In place of the 3D equation  (\ref{eq3}), a quasi-2D model appropriate for the 
quasi-2D RDB  solitons of large spatial extension  is very economic and convenient 
from a computational point of view, specially for real-time dynamics. 
The system is assumed to be in the ground state $\phi(y)=\exp (-y^2/2)/(\pi )^{1/4}$
of the axial  trap and the wave function can be written as 
$\phi({\bf r},t)=\phi(y) \phi_{2D}({\vec \rho},t)$, where ${\vec \rho} \equiv \{x,z\}$
and $ \phi_{2D}({\vec \rho},t)$ is an effective wave function in the $x-z$ plane. Using this ansatz in (\ref{eq3}), the $y$ dependence can be integrated out to obtain the following effective 2D equation \cite{laser2D,dipolarCPC}
\begin{eqnarray}&
{ i} \frac{\partial \phi_{2D}({\vec \rho},t)}{\partial t}=
\biggr[ - \frac{\nabla_\rho^2}{2}+ \frac{g}{\sqrt{2\pi}}|\phi_{2D}
({\vec \rho},t)|^2 +\frac{4\pi g_{\mathrm{dd}}}{3\sqrt{2\pi}} \label{eq5}
\nonumber \\
&\times \int \frac{d{\bf k_\rho}}{(2\pi)^2}
e^{-i{\bf k_\rho}\cdot {\vec \rho}} n({\bf k_\rho},t)j_{2D}(\xi) \biggr] \phi_{2D}({\vec \rho},t),     \\
&n({\bf k_\rho},t)= \int e^{i{\bf k_\rho}\cdot {{\vec \rho}}}
|\phi_{2D}({\vec \rho},t)|^2
d
{\vec \rho},\quad  {\bf k_\rho} \equiv (k_x,k_y), \\&
 j_{2D}(\xi)= -1+3\sqrt \pi \frac{k_z^2}{2\xi} e^{\xi^2}[1-\mathrm{erf}(\xi)], \quad  \xi=\frac{k_\rho}{\sqrt 2}.
\end{eqnarray} 
In this case $\int 
|\phi_{2D}({\vec \rho,t})|^2 d {\vec \rho}=1$.

\section{Numerical Results}

We solve the 3D equation (\ref{eq3})  or the 2D equation (\ref{eq5})
by the split-step 
Crank-Nicolson discretization scheme using  real-time propagation
  in 3D or 2D Cartesian coordinates, respectively,
using a space step of 0.1 $\sim$ 0.2
and a time step of  
 $0.0002 \sim 0.002$ in real-time simulation \cite{dipolarCPC,CPC}. 
A small time step is employed in the real-time propagation to obtain reliable and accurate results. The dipolar potential term is treated by a Fourier transformation to the momentum space using a convolution rule \cite{Santos01}.

\begin{figure}[!t]

\begin{center}
\includegraphics[width=\linewidth,clip]{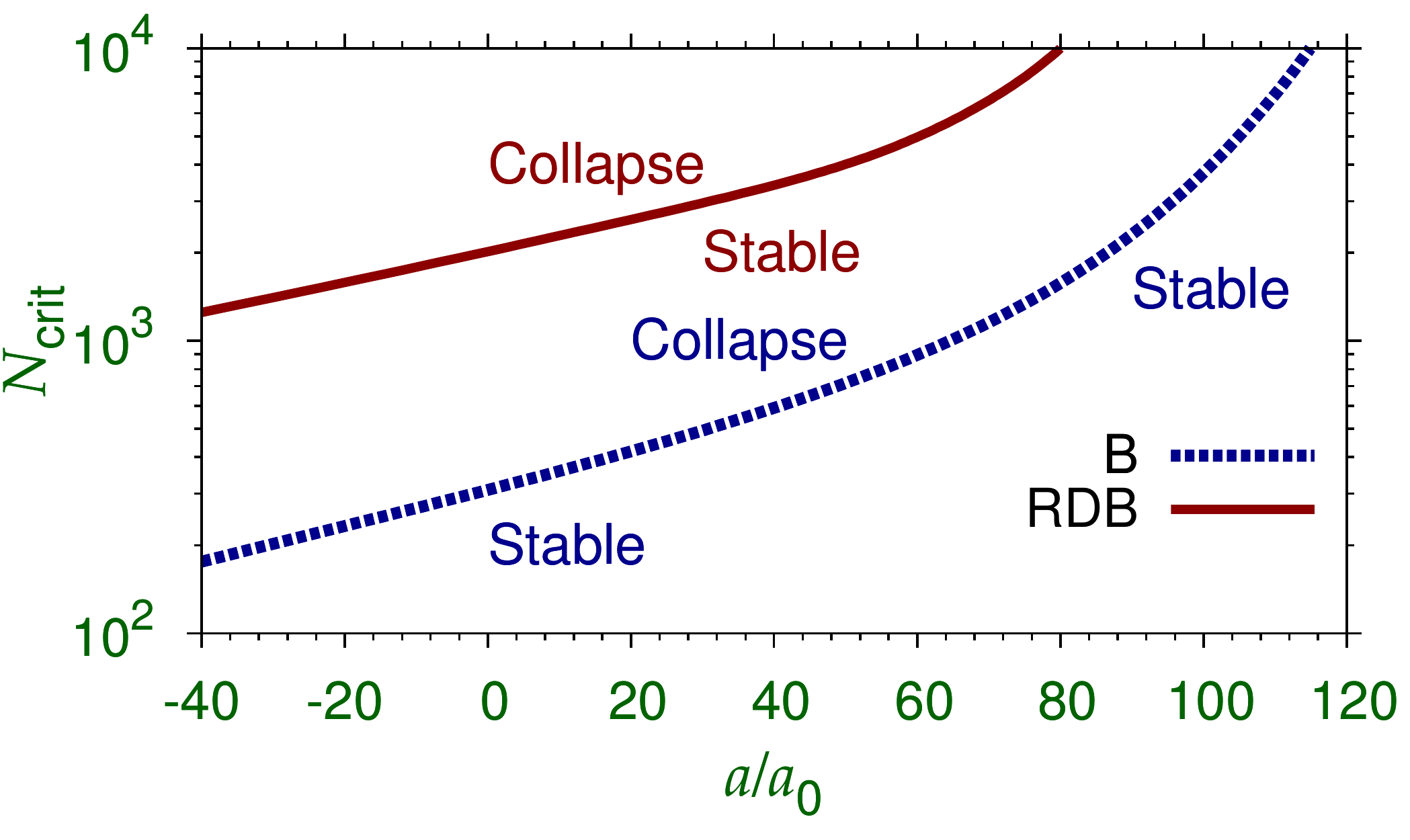}

\caption{ (Color online) Critical number of $^{164}$Dy  atoms $N_{\mathrm{crit}}$ for the formation of a stable 2D bright (B) and RDB soliton as obtained from a solution of 
   (\ref{eq5}). The oscillator length $l\equiv\sqrt{\hbar/mw}=1$ $\mu$m and the dipolar length $a_{\mathrm{dd}}=132.7 a_0$. The soliton collapses for for a number of atoms $N> N_{\mathrm{crit}}$.
}\label{fig1} \end{center}

\end{figure}

In 2D, the simulation is started with the normalized excited state 
\begin{eqnarray}\label{pqr2}
\phi_{2D}({\vec \rho})=\sqrt{\frac{\kappa}{\pi}}[\kappa(x^2+z^2)-1]e^{-\kappa(x^2+z^2)/2}
\end{eqnarray}
with eigenvalue $3\kappa$ of   (\ref{eq5}) with 
$g=g_{\mathrm{dd}}=0$ and  harmonic trap $\kappa^2(x^2+ z^2)/2$.  
In 3D, the  excited state  
\begin{eqnarray}\label{pqr3}
\phi({\bf r})=\sqrt{\frac{\kappa}{{\pi^{3/2}}}}[\kappa(x^2+z^2)-1]e^{-\kappa(x^2+z^2)/2-y^2/2}
\end{eqnarray}
with eigenvalue $3\kappa+1/2$ of    (\ref{eq3}) 
with  $g=g_{\mathrm{dd}}=0$ and harmonic trap $\kappa^2(x^2+ z^2)/2+y^2/2$  is used.  States (\ref{pqr2}) and (\ref{pqr3})  have a circular notch at $\rho\equiv \sqrt{x^2+z^2}=\sqrt{1/\kappa}$. The stationary profile of the    quasi-2D RDB solitons can be obtained by real-time simulation. The 2D dipolar RDB solitons are excited states of 2D bright solitons  and are not accessible by imaginary-time propagation which is designed to converge to the ground states of    (\ref{eq3}) and (\ref{eq5}). 
The real-time simulation is started with a small $\kappa (\sim 0.0025)$ 
corresponding to a circular notch at $\rho \sim 20$
and with $g=g_{\mathrm{dd}}=0$. The final converged result is independent of the initial choice of $\kappa$, but this initial choice leads to a quick convergence of the result. 
In the course of simulation 
the nonlinearities are increased slowly to the desired values and the    trap is slowly reduced to zero. In this way the real-time simulation passes through eigenstates of the GP equation until we finally reach the stable RDB soliton with a circular notch as our final state.

\begin{figure}[!t]

\begin{center}
\includegraphics[width=\linewidth,clip]{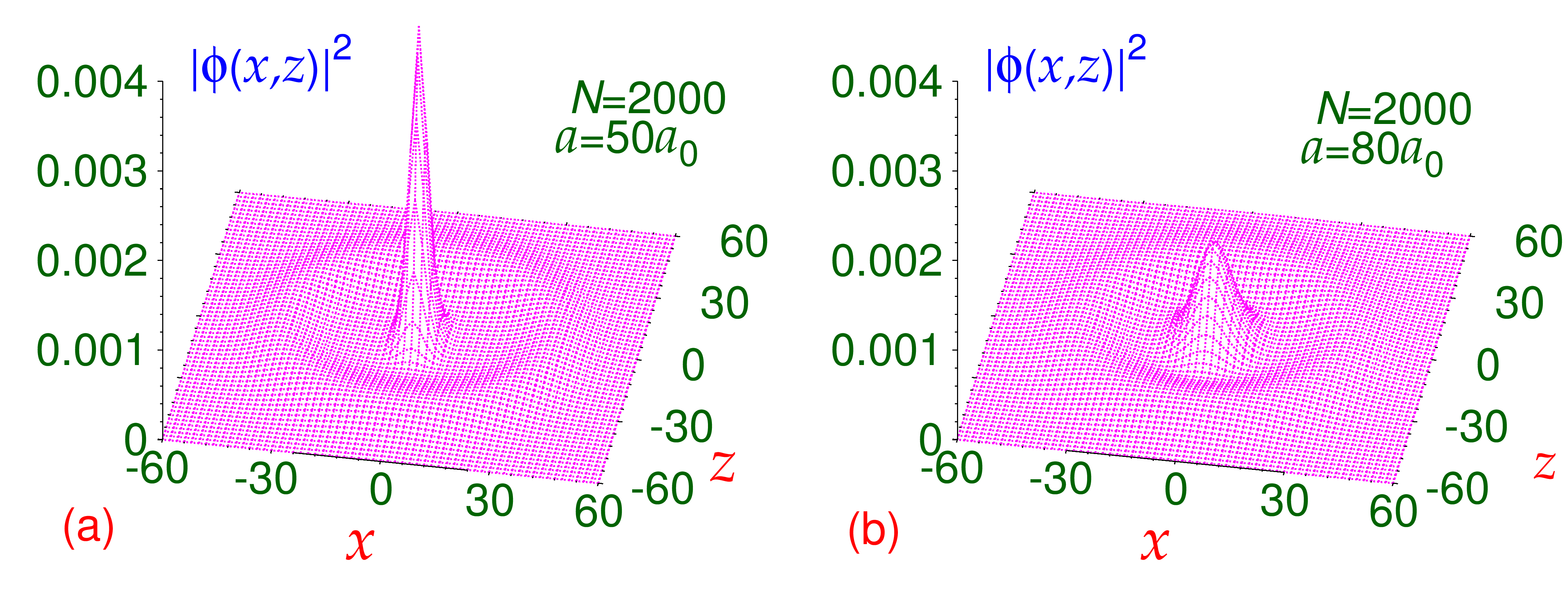}
\includegraphics[width=\linewidth,clip]{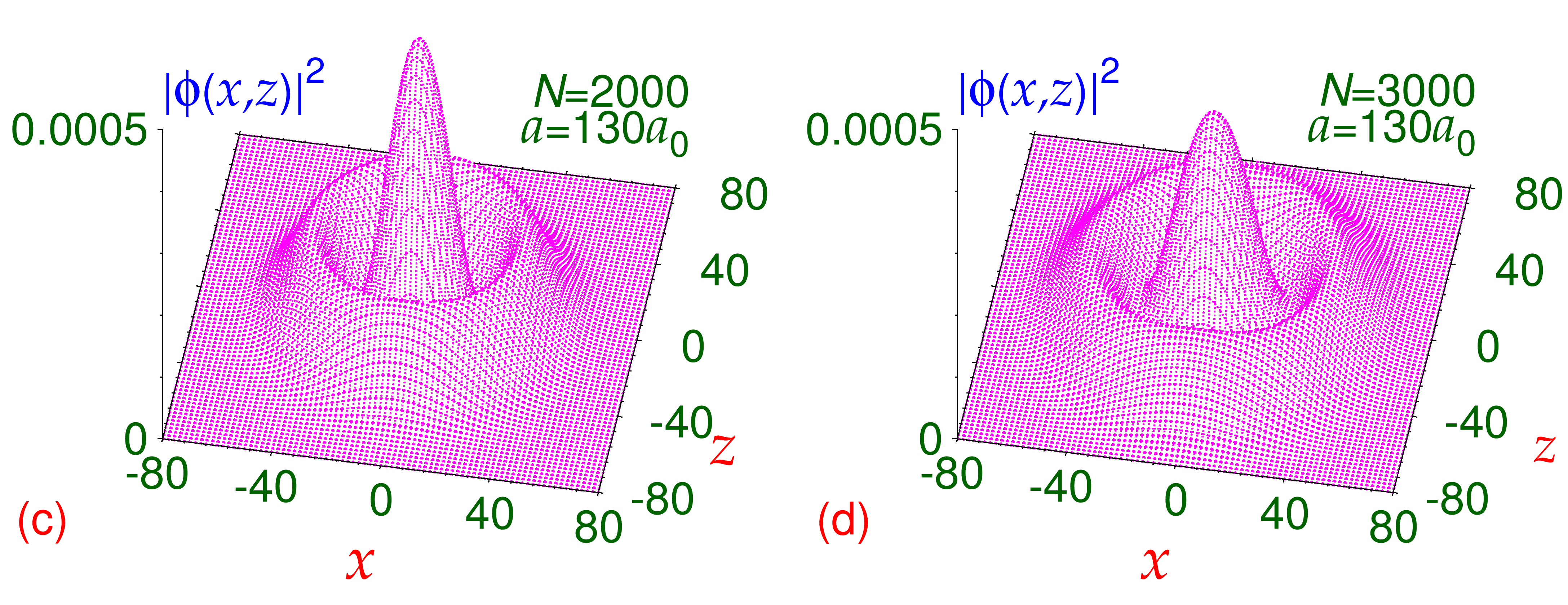}

\caption{ (Color online) The density $|\phi({\vec \rho})|^2\equiv |\phi(x,z)|^2$  of the 2D dipolar RDB solitons for (a) $N=2000, a=50a_0$,
(b) $N=2000, a=80a_0$, (c)  $N=2000, a=130a_0$, and (d) $N=3000, a=130a_0$ as obtained from   (\ref{eq5}). The densities are normalized as $\int dx dz |\phi(x,z)|^2=1$ and expressed in units of $l^{-2}$ and the length in units of $l$.
The length scale $l=1$ $\mu$m and the dipolar length $a_{\mathrm{dd}}=132.7 a_0$.
}\label{fig2} \end{center}

\end{figure}

\begin{figure}[!t]

\begin{center}
\includegraphics[trim = 0mm 2mm 0mm 2mm, clip,width=.3\linewidth]{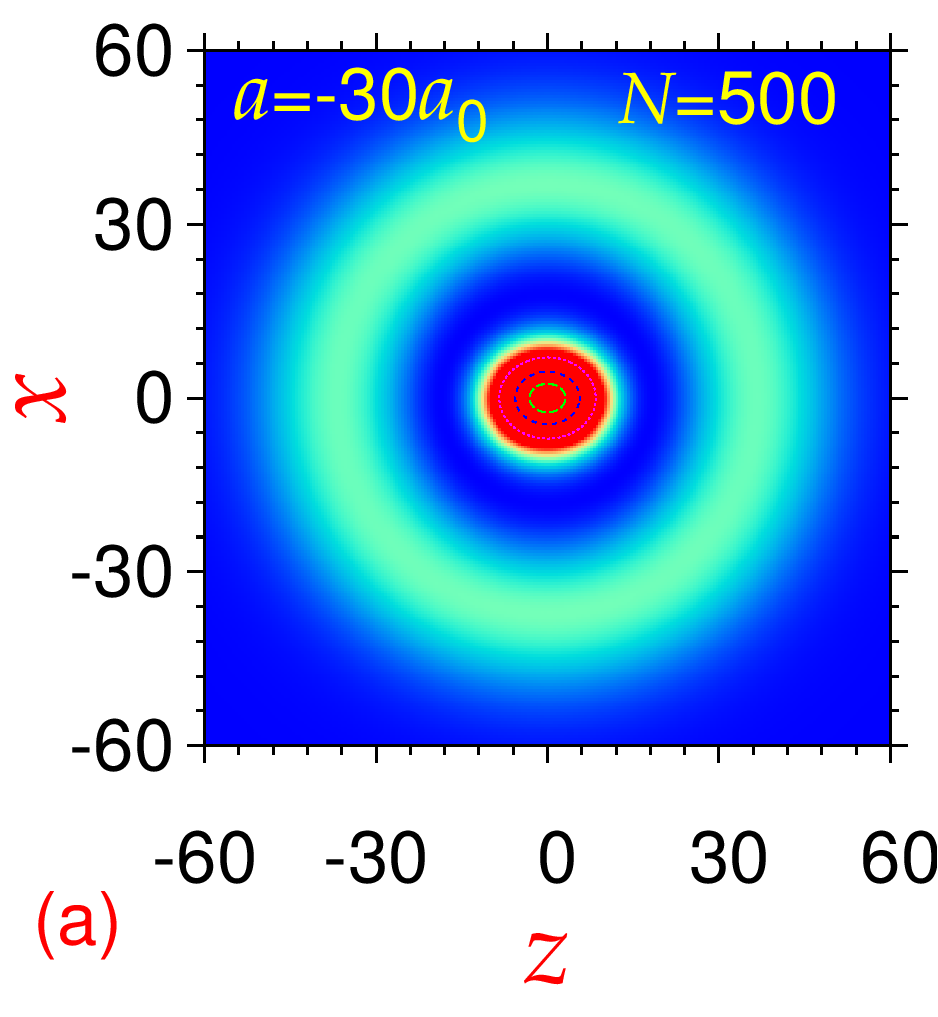}
 \includegraphics[trim = 0mm 2mm 0mm 2mm,  clip,width=.3\linewidth]{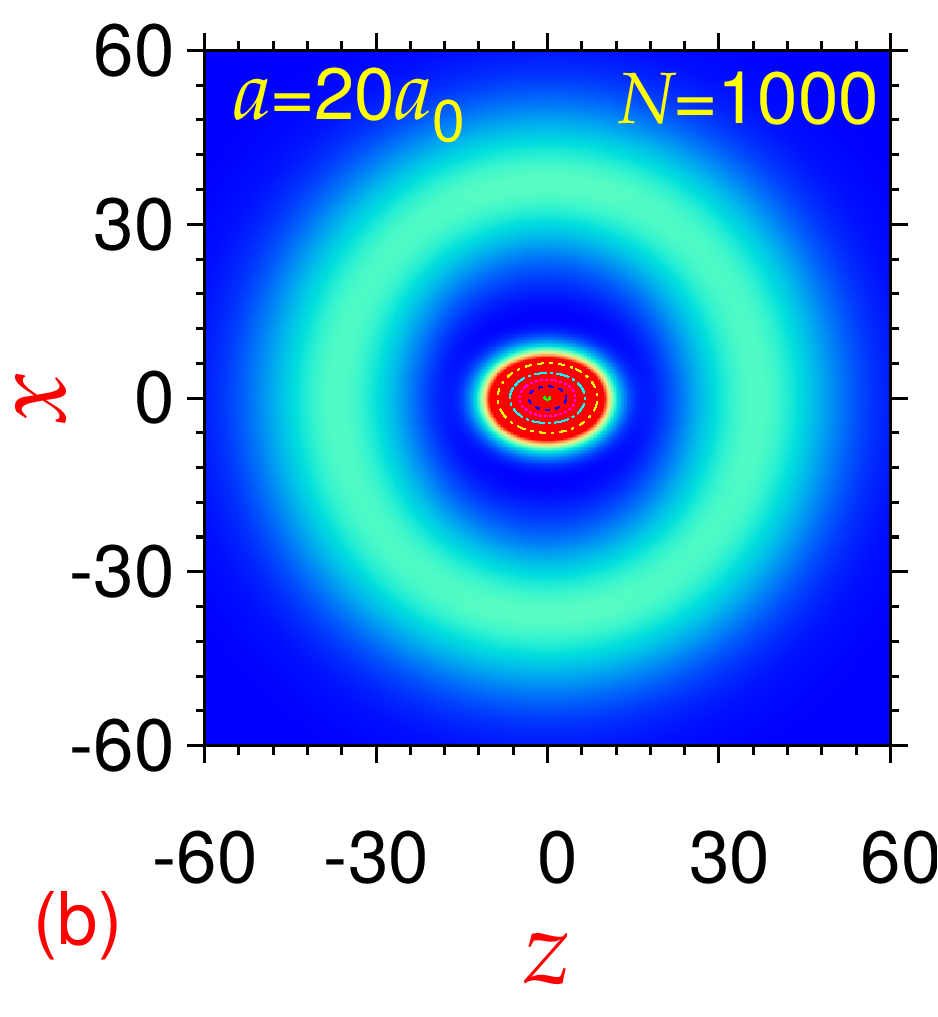}
\includegraphics[trim = 0mm 2mm 0mm 2mm,  clip,width=.3\linewidth]{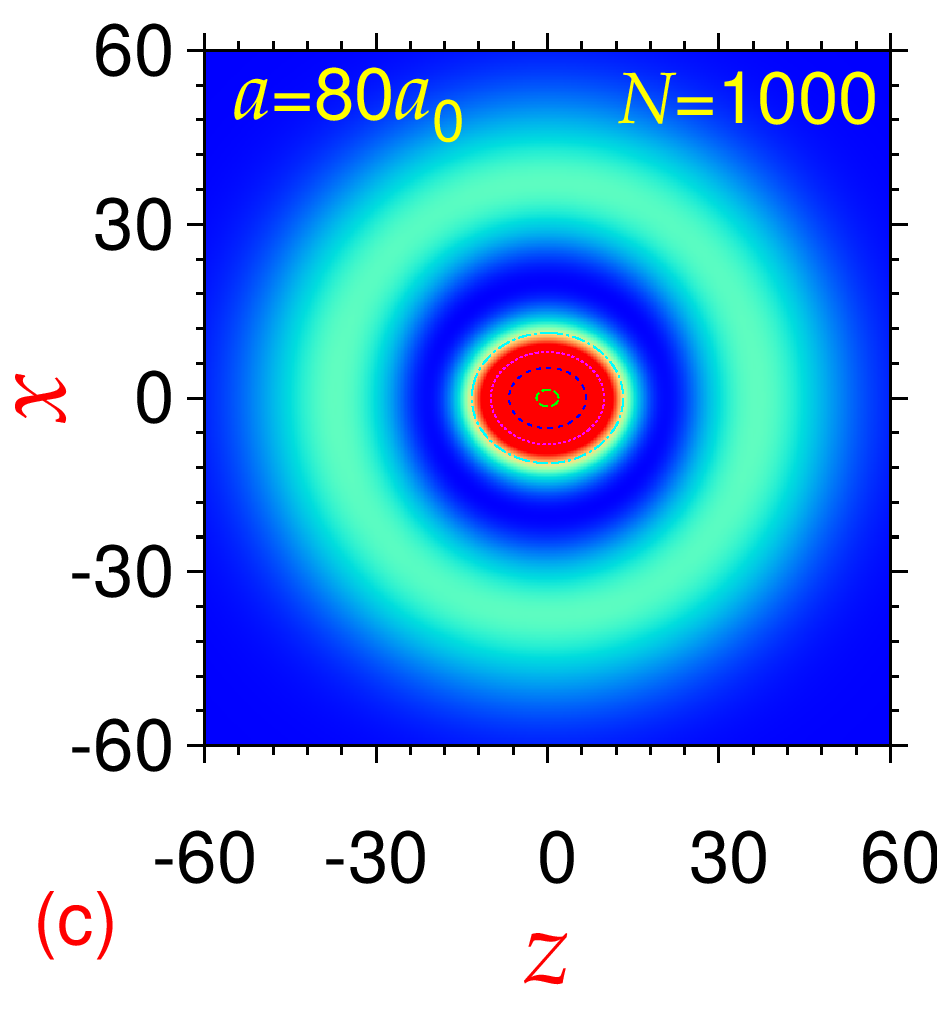}
 \includegraphics[trim = 0mm 2mm 0mm 2mm,  clip,width=.3\linewidth]{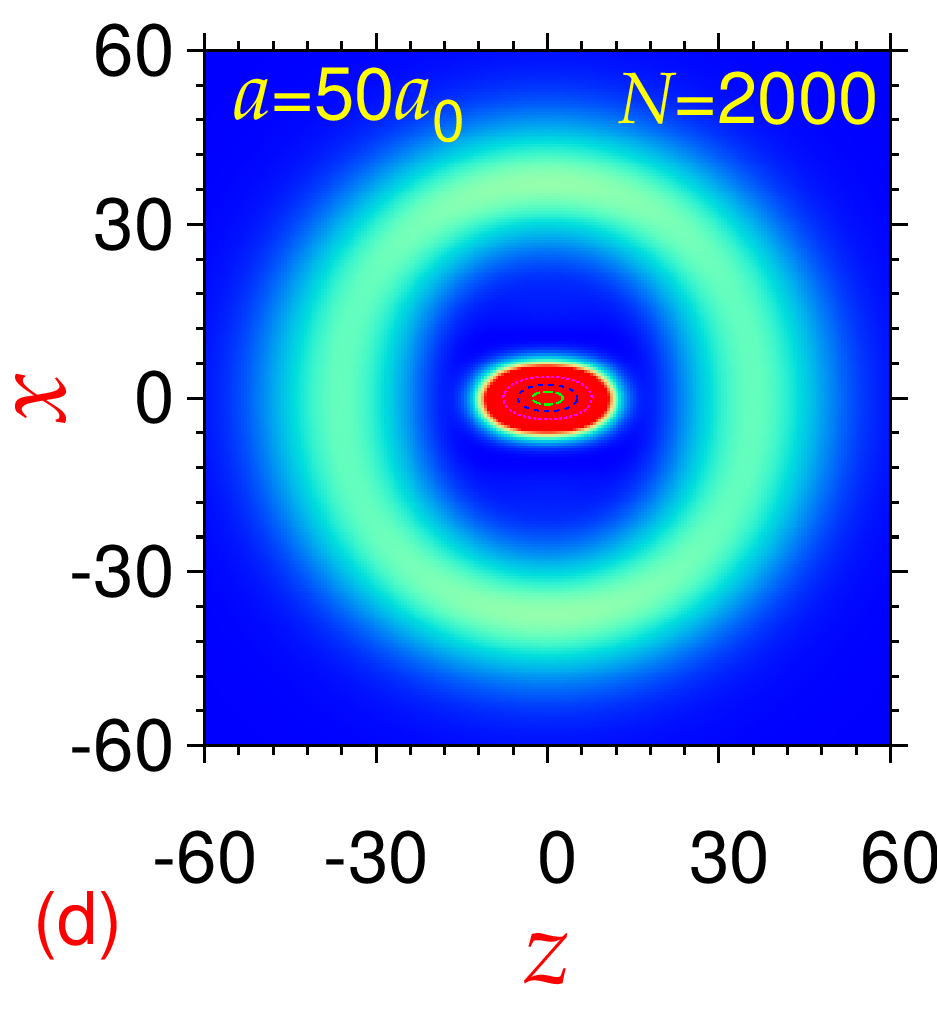}
 \includegraphics[trim = 0mm 2mm 0mm 2mm,  clip,width=.3\linewidth]{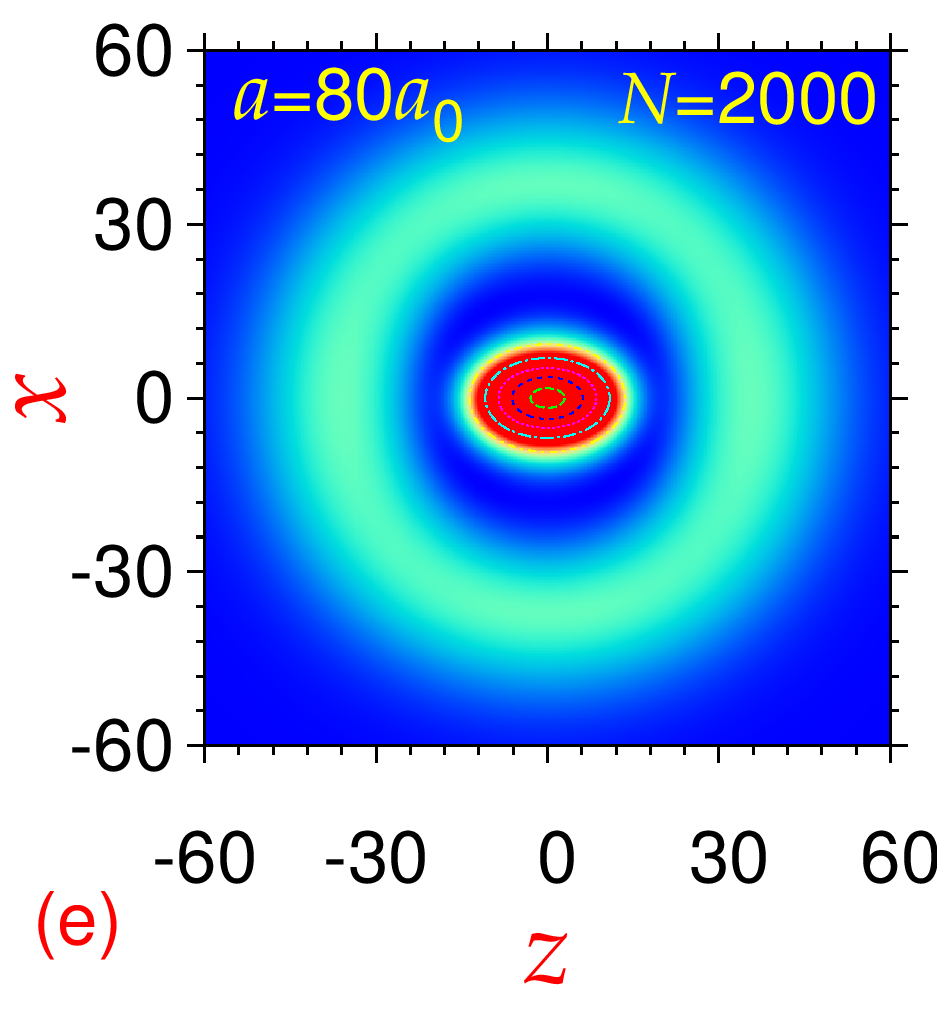}
\includegraphics[trim = 0mm 2mm 0mm 2mm,  clip,width=.3\linewidth]{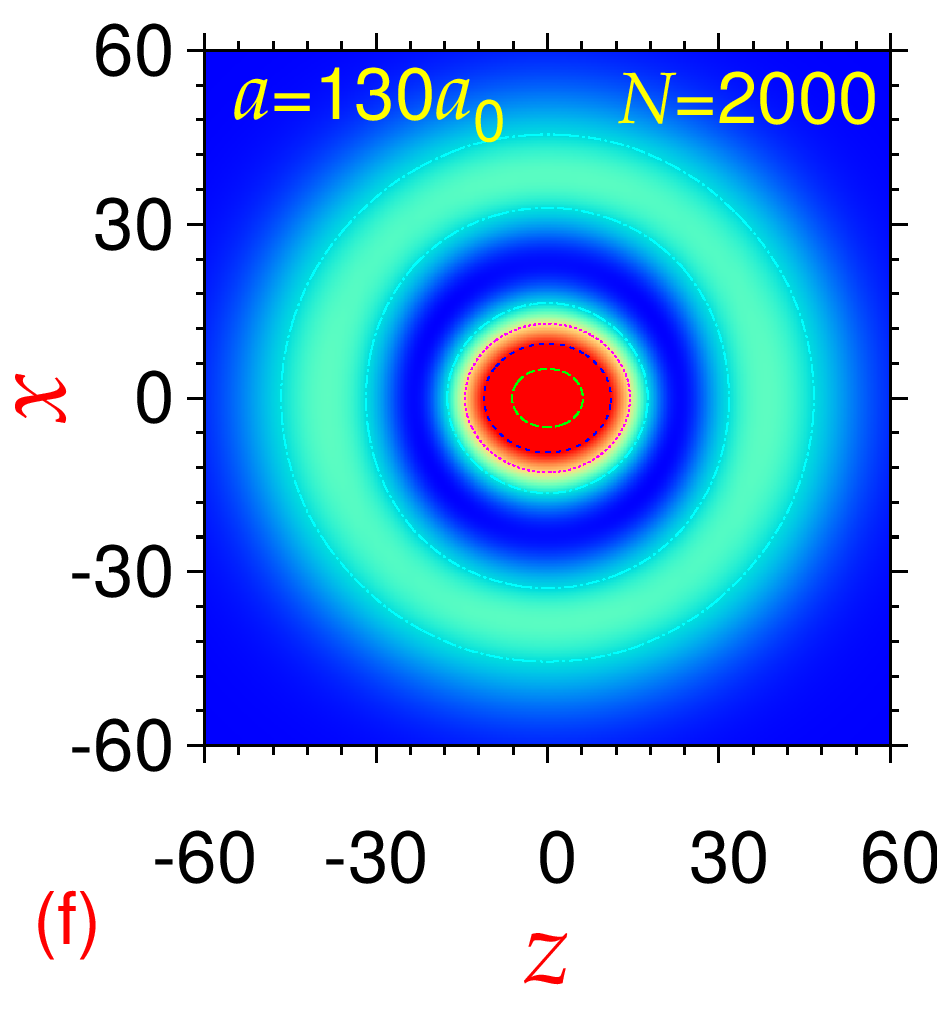}
\includegraphics[trim = 0mm 1mm 0mm 2mm,  clip,width=.3\linewidth]{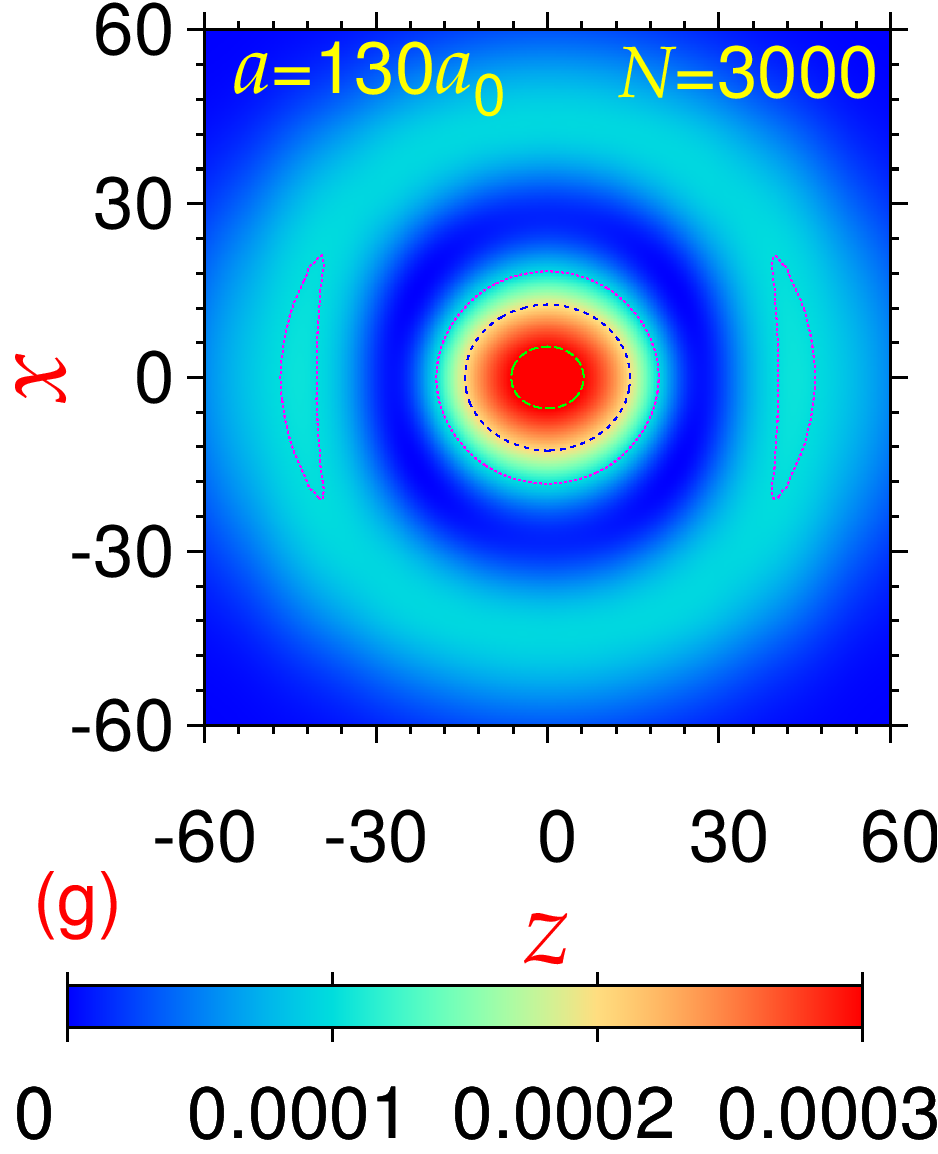}
\includegraphics[trim = 0mm 1mm 0mm 2mm, clip,width=.3\linewidth]{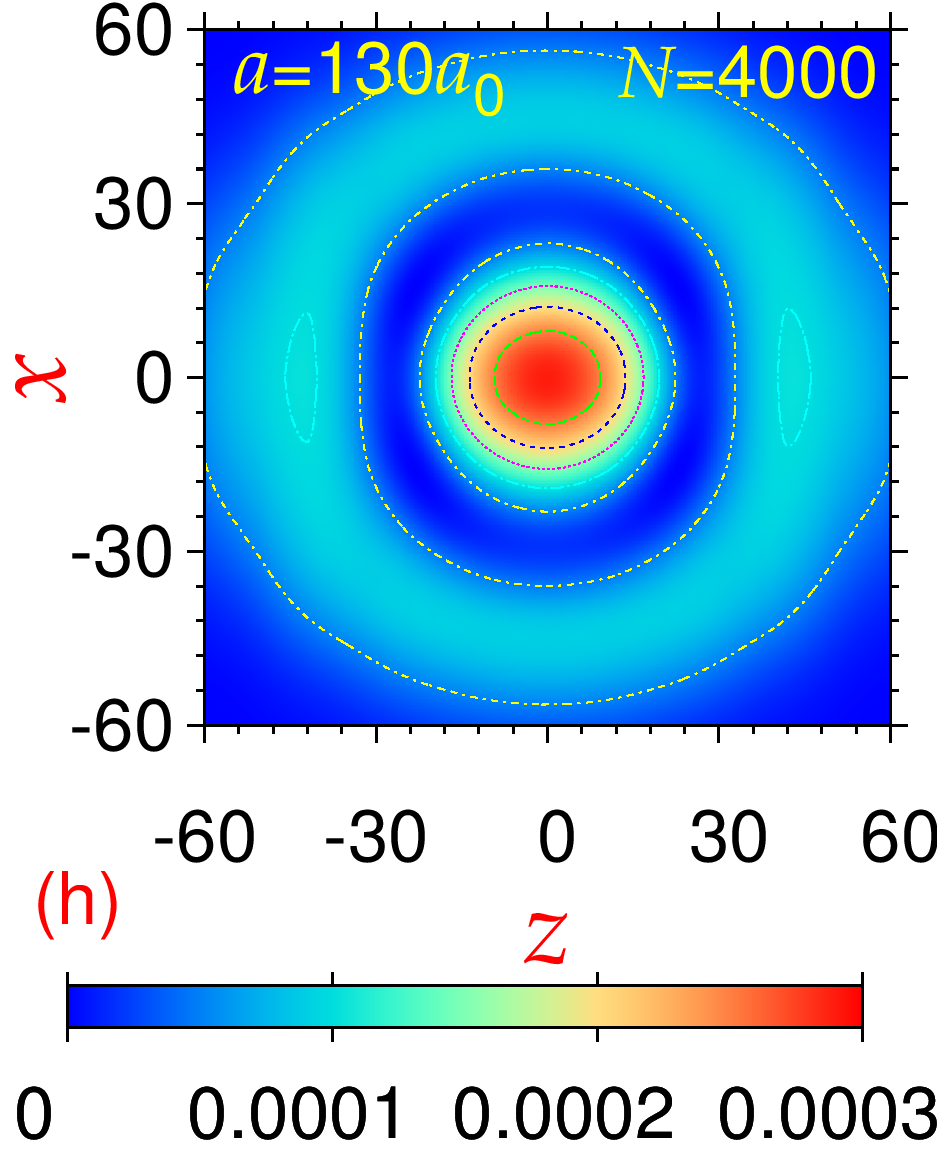}

\caption{ (Color online)  The contour plot of normalized  density $|\phi({\vec \rho})|^2\equiv |\phi(x,z)|^2$ of the 2D RDB solitons from a solution of  (\ref{eq5}) for (a) $N=500, a=-30a_0$,
(b) $N=1000, a=20a_0$, (c)  $N=1000, a=80a_0$,  (d) $N=2000, a=50a_0$,
(e) $N=2000, a=80a_0$,
(f) $N=2000, a=130a_0$, (g)  $N=3000, a=130a_0$, and (h) $N=4000, a=130a_0$.  The densities are   expressed in units of $l^{-2}$ and the length in units of $l$
and the length scale $l=1$ $\mu$m.
}\label{fig3} \end{center}

\end{figure}

In this study we consider  2D  RDB solitons in 
a BEC of $^{164}$Dy atoms with magnetic moment $\mu = 10 \mu_B$ \cite{ExpDy}, where $\mu_B$ 
is the Bohr magneton. 
The   $^{164}$Dy atoms  have the 
largest magnetic moment among those used in dipolar BEC experiments and a large dipole
moment is fundamental for achieving   2D RDB    solitons 
with a large number of atoms.   
The dipolar length in this case is $a_{\mathrm{dd}}= \mu_0\mu^2 m/(12 \pi \hbar^2)= 132.7 a_0$ where $a_0$ is the Bohr radius.  
We take the  length scale $l =1$ $\mu$m, corresponding  to a harmonic trap frequency of  $\omega = 2\pi \times 61.6$ Hz along $y$ direction and, consequently,  the time scale $t_0= 2.6 $ ms.

\begin{figure}[!t]

\begin{center}
\includegraphics[trim = 2mm 1mm 1mm 5mm, clip, width=.48\linewidth,clip]{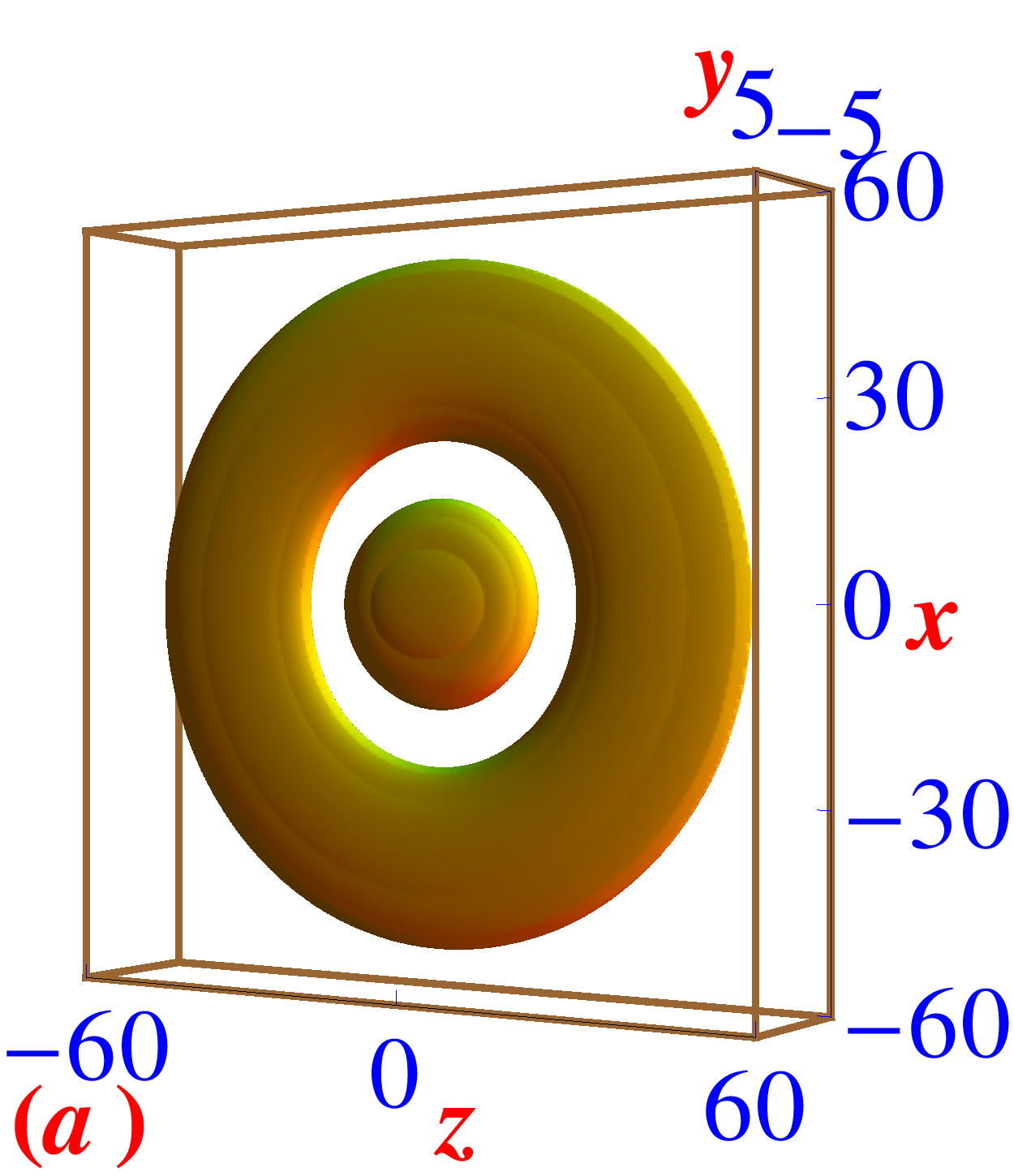}
\includegraphics[trim = 1mm 0mm 0mm 0mm, clip, width=.48\linewidth,clip]{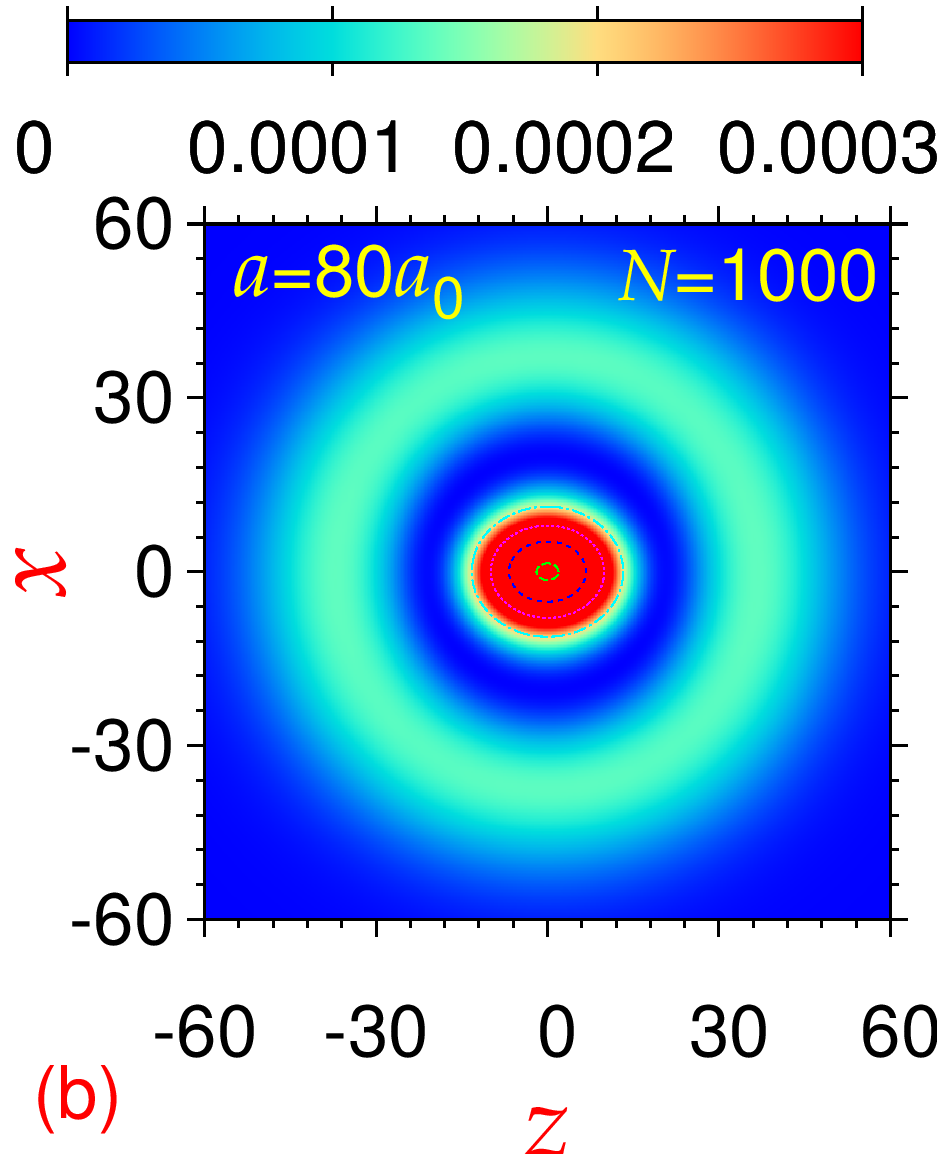}

\caption{ (Color online) (a) Isodensity profile of the 2D RDB soliton obtained from a solution of the 3D GP equation (\ref{eq3}) for $N=1000$ and $a=80a_0$. The 3D contour is set at 0.00002. (b) The contour plot of the quasi-2D density $|\phi_{2D}(x,z)|^2=\int dy|\phi({\bf r})|^2$ of the 3D calculation. 
}\label{fig4} \end{center}

\end{figure}

We study the domain of the appearance  of the  2D bright   solitons  of   (\ref{eq5}).
These solitons for a specific   scattering length can exist 
for the number of atoms $N$ below a critical number $N_{\mathrm{crit}}$ beyond which the system collapses \cite{jbohn}. In figure \ref{fig1} we plot this critical 
value $N_{\mathrm{crit}}$ versus $a/a_0$ from imaginary-time simulation.    
The   2D dipolar RDB solitons, as obtained by real-time simulation,    have a much  larger spatial extension compared to the bright solitons and can accommodate a larger number of atoms
as can be seen from figure \ref{fig1}.
 For $N>N_{\mathrm{crit}} $,  the soliton collapses 
due to an excess of dipolar attraction. In the stable region there is a balance between attraction and repulsion to form the bright soliton. 
The number of atoms in the 2D dipolar bright and RDB solitons   could be quite large and will be of  experimental interest. 
The size of quasi-1D nondipolar solitons   is usually quite small and   accommodates only a small
number of atoms.

Next we present the profile of the 2D RDB solitons as obtained from   (\ref{eq5})
by real-time propagation for different number of  $^{164}$Dy atoms and different scattering lengths. The scattering length can be adjusted to a desired value by the magnetic \cite{fesh}
and optical \cite{optfesh}
  Feshbach resonance technique. The densities of the different RDB solitons for (a) $N=2000, a=50a_0$,
(b) $N=2000, a=80a_0$,
(c) $N=2000, a=130a_0$, and  (d)  $N=3000, a=130a_0$,  
are shown in figure \ref{fig2}.
A Gaussian input wave function converges to the   2D bright soliton studied in Ref. \cite{2D}.
  The input functions (\ref{pqr2}) and (\ref{pqr3}) in   (\ref{eq3}) and (\ref{eq5}), respectively, with a circular notch at $\rho =1/\sqrt\kappa$  representing a radial excitation, converge to the RDB soliton profiles shown in figure \ref{fig2}.  Due to the 
anisotropic dipolar 
interaction, the densities of the RDB solitons are anisotropic in the $x-z$ plane  when the dipolar interaction dominates the contact interaction ($a_{\mathrm{dd}}\gg a$). 
Although the 
notch in the density of the RDB soliton can clearly be seen in plots of figure \ref{fig2}, 
the anisotropic shapes of the densities can be clearly seen in contour plots of densities and 
this is why we consider the contour plots of density in the following.

In figure \ref{fig3} we display the contour plots of the density of 2D RDB solitons  
for different number $N$ of $^{164}$Dy atoms and   scattering lengths $a$. From 
figures \ref{fig2} and \ref{fig3} and other calculations not presented here, several aspects of the density profiles of the RDB solitons are clear.  For a small number of atoms ($N<1000$) the dipolar nonlinearity is small and consequently the density profile possesses approximate circular symmetry in the $x-z$ plane. Deviation from circular symmetry occurs for a larger number of atoms and for $a_{\mathrm{dd}}\gg a$ when the dipolar interaction plays a dominating role in the formation of the RDB solitons. Comparing figures \ref{fig3}(c) and (e) for 1000 and 2000 atoms with scattering length $a=80a_0$,  we see that the anisotropy is larger for a 
larger number of atoms.   Comparing figures \ref{fig3}(d), (e) and (f) for $a=50a_0, 80a_0$ and $130a_0$, respectively, for $N=2000$ we find that the anisotropy has reduced with an increase of the scattering length $a$ from   $a=50a_0$ to  $130a_0$. 
For $a=130a_0 \approx a_{\mathrm{dd}}=132.7a_0$, the anisotropic dipolar interaction plays a less 
dominating role in the formation of the RDB soliton and hence results in a reduced anisotropy in figures \ref{fig3}(f), (g) and (h) for $N=2000, 3000,$ and 4000.

\begin{figure}[!t]

\begin{center}
\includegraphics[width=.6\linewidth,clip]{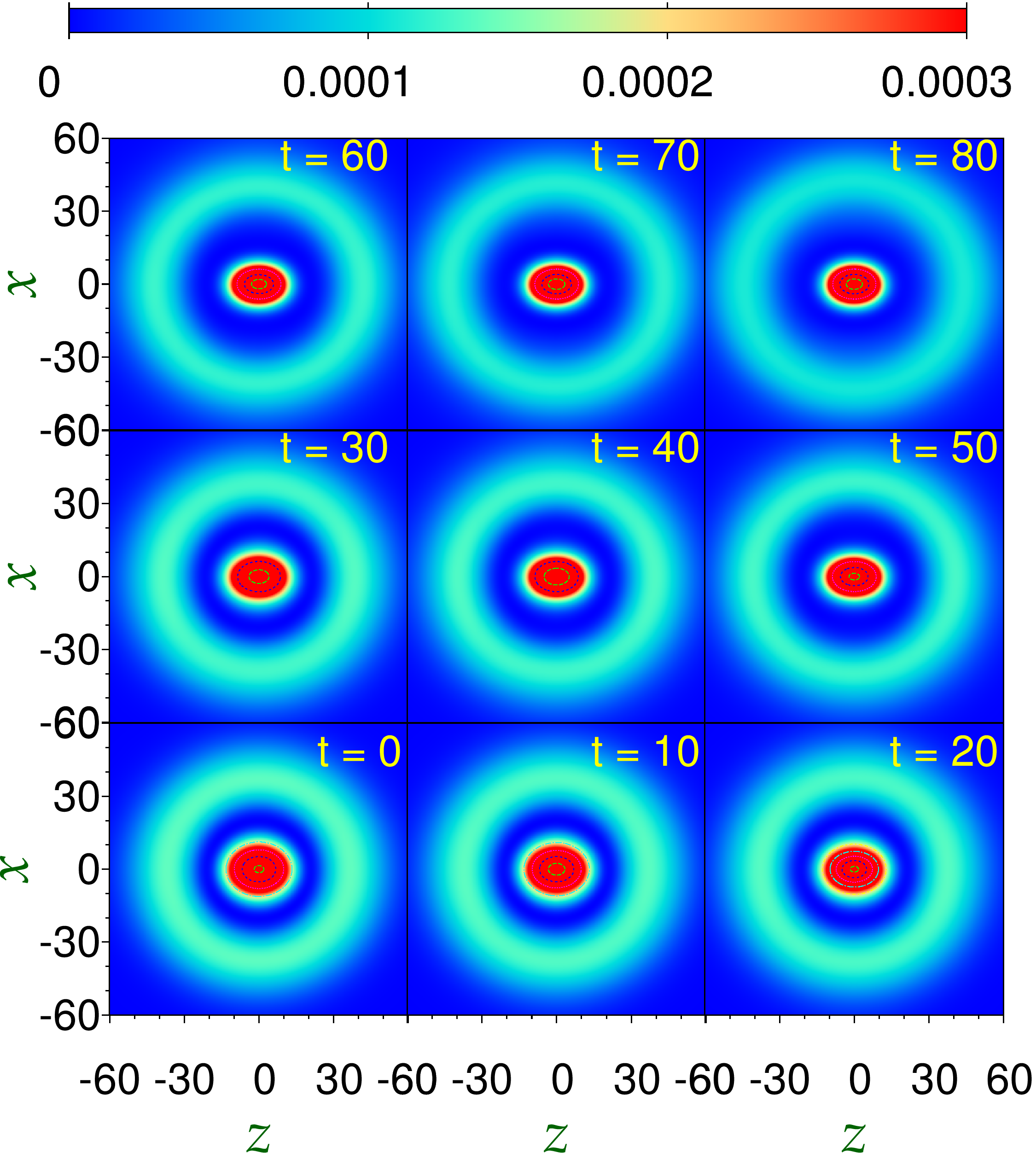}

\caption{ (Color online) Contour plot of density at different times $t$ as obtained
from   (\ref{eq5}) 
illustrating the 
dynamics of a RDB soliton of 1000 $^{164}$Dy atoms of scattering 
length $a=80a_0$ when at $t=0$ the scattering length is suddenly changed to $a=20a_0$. 
The simulation is started with the wave function  of the   RDB soliton exhibited in 
figure \ref{fig3}(c).  
}\label{fig5} \end{center}

\end{figure}

The results reported in this paper are performed with the quasi-2D GP equation (\ref{eq5}). Although this seems very reasonable in the presence of a strong trap in the transverse $y$ direction, it is worthwhile to compare these densities with those obtained from a solution of the 3D GP equation (\ref{eq3}) and we do this in the following. We calculated the density of several of the RDB solitons illustrated in figure \ref{fig3} using the 3D GP  equation (\ref{eq5}). The quasi-2D densities obtained from the 2D and 3D GP equations are very similar in all cases. In figure \ref{fig4}(a) we show the 3D Isodensity profile of the RDB soliton obtained from  (\ref{eq3}) for $N=1000$ and $a=80a_0$. The quasi-2D density $|\phi_{2D}(x,z)|^2=\int dy|\phi({\bf r})|^2$ obtained from the same calculation is shown in figure 
\ref{fig4}(b). This quasi-2D density is very similar to the same presented in figure \ref{fig3}(c) from a solution of the quasi-2D GP equation (\ref{eq5}), which validates the use of this 
equation.

\begin{figure}[!t]

\begin{center}
\includegraphics[trim = 3mm 0mm 1.5mm 1mm, clip,width=.3\linewidth,clip]{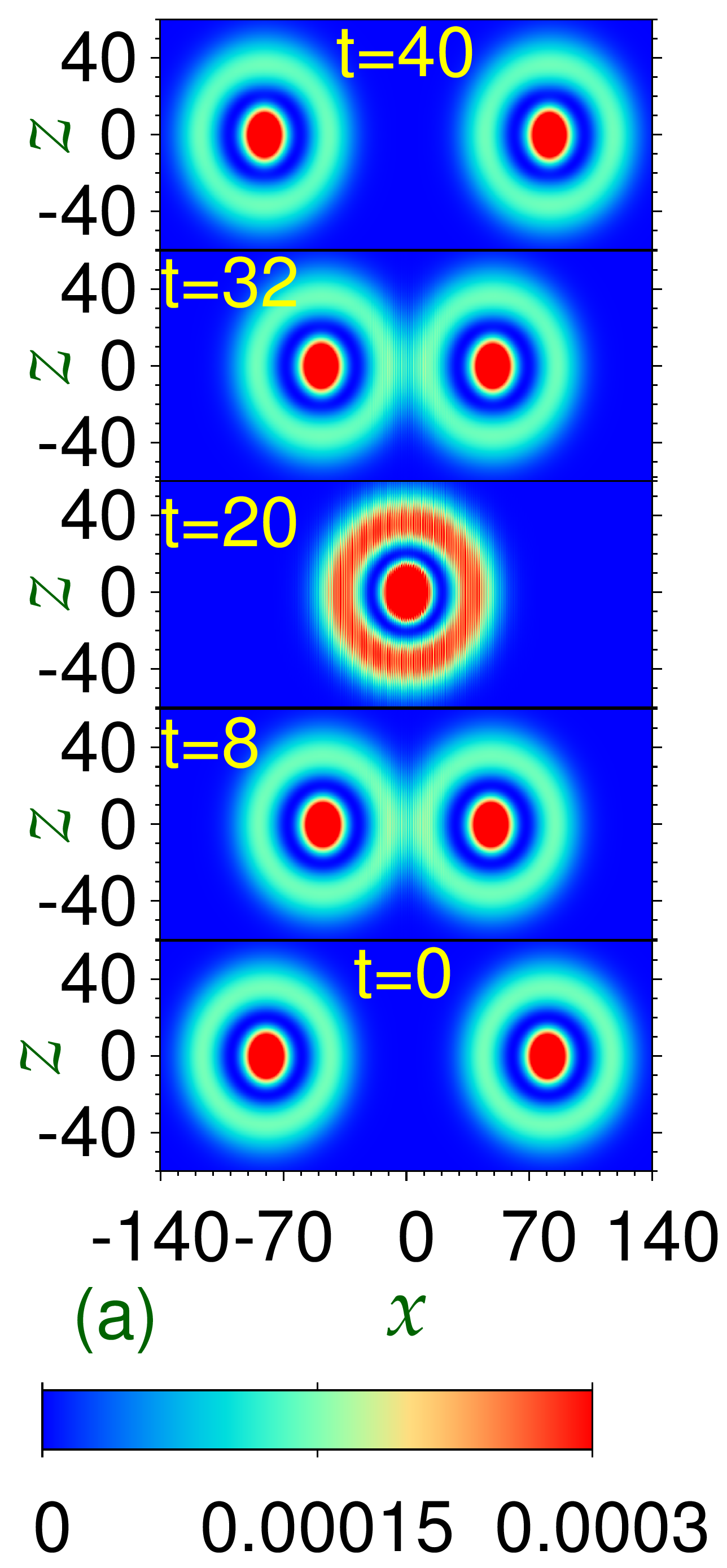}
\includegraphics[trim = 3mm 0mm 1.5mm 1mm, clip,width=.3\linewidth,clip]{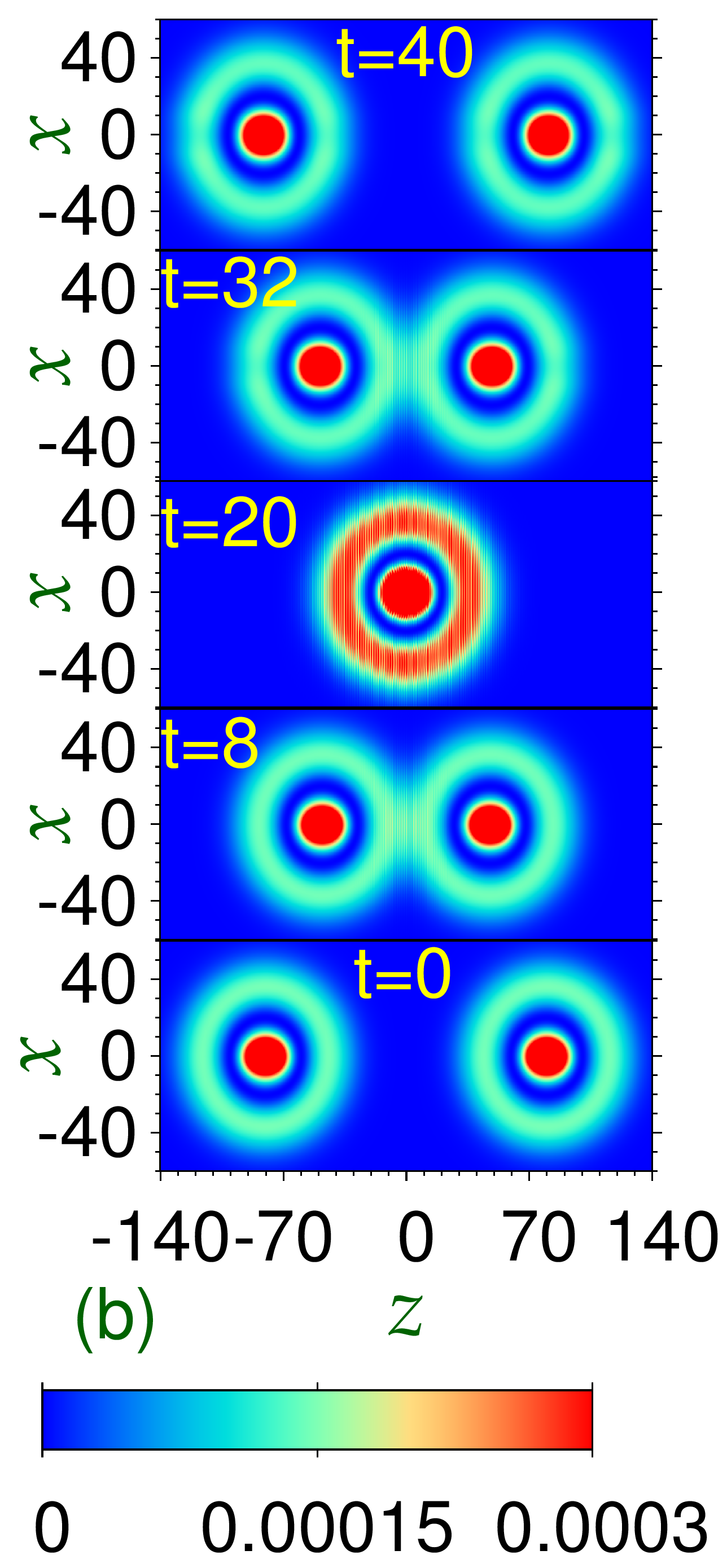}

\caption{ (Color online) Contour plot of density at different times $t$ 
as obtained from   (\ref{eq5}) illustrating the 
quasi elastic collision of two RDB solitons of 1000 $^{164}$Dy atoms of scattering length $a=80a_0$
moving in opposite directions along (a) $x$ and (b) $y$ axes, respectively, with a dimensionless velocity of 4 units.  With the length  scale $l=1$ $\mu$m the velocity of each 
RDB soliton is about 1.5 mm/s.
}\label{fig6} \end{center}

\end{figure}

The  2D RDB solitons presented in figures \ref{fig2} and \ref{fig3}
are stable and  robust as tested under real-time propagation  with a reasonable perturbation in the parameters.  The robustness comes from the large contact repulsion   which  inhibits collapse. Also, the predominant long-range 
dipolar attraction in the $x-z$ plane prevents the leakage of the atoms to infinity. In a nondipolar quasi-1D BEC soliton, the contact attraction alone provides the binding and there is no repulsion to stop the collapse.  We provide below two numerical tests for the stability 
of a RDB soliton.

First we test the stability of  a RDB soliton under a sudden change of the scattering length.
We perform real-time simulation,  with the wave function of the RDB soliton of figure \ref{fig3}(c) as the initial state, when at time $t=0$ the scattering length is suddenly changed from $a=80a_0$ to $20a_0$. The contour plots of density at different times is exhibited in Fig, \ref{fig5}.
Although the notch in the density remains intact, at large times the RDB soliton adjusts to the shape with 
$a=20a_0$ shown in figure \ref{fig3}(b). The anisotropy has increased with time due to an increased contribution of dipolar interaction as the  contact interaction is reduced   at $t=0$.

\begin{figure}[!t]

\begin{center}
\includegraphics[width=.6\linewidth,clip]{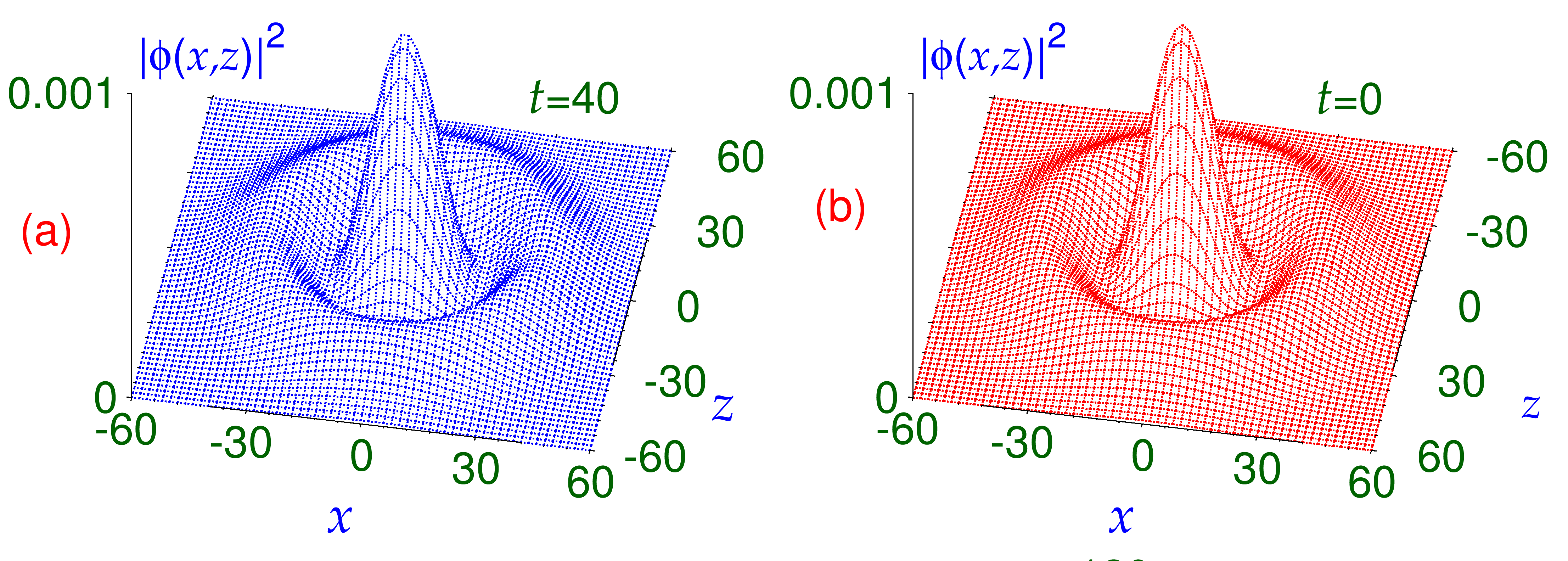}

\caption{ (Color online) The (a) final ($t=40$) and (b) initial ($t=0$) profiles of a 
RDB soliton of 1000 atoms and $a=80a_0$  undergoing collision in figure \ref{fig6}(b) moving along the $z$ axis. 
}\label{fig7} \end{center}

\end{figure}

 A stringent test of the robustness 
of these 2D RDB solitons is provided in their behavior under head-on collision. Like the 1D dipolar solitons \cite{1D}, the collision of  the RDB solitons is expected to be quasi elastic with the solitons emerging with little deformation after collision  at medium velocities.   Only the collision between two integrable 1D solitons is known to be perfectly elastic at all velocities \cite{rmp}.  We consider a head-on collision between two  2D RDB solitons of figure \ref{fig3}(c) moving in opposite directions along both $x$ and $z$ axes.     The collision 
dynamics of two such solitons as generated from a real-time simulation of Eq. (\ref{eq5})
is
shown in figure \ref{fig6}(a) and (b), respectively. 
The initial velocities of the two solitons  are attributed by multiplying the initial wave functions of the two solitons by  phase factors
$\exp (\pm iv_x x), v_x=20$.  The two  2D dipolar RDB solitons are initially placed at $x\equiv x_0=\pm 80$ and the real-time simulation started. The two solitons move in opposite directions and suffer a head-on collision. The collision dynamics for velocity along $x$ direction  is illustrated by plotting  the density of the system at different times as shown in figure \ref{fig6}(a). 
  The dimensionless velocity of each   soliton is about $4$, which corresponds to about 1.5 mm/s using scales $l=1$ $\mu$m and $t_0=2.6$ ms. In figure \ref{fig6}(b) we show the collision dynamics for velocity along $z$ direction.
The smooth density profiles of the dynamics presented in figure \ref{fig6} illustrates the quasi elastic nature of the dynamics. In figure \ref{fig7}(a) we plot the final density of one
of the solitons after collision at $t=40$ shown in figure \ref{fig6}(b) and compare it with the initial density before collision shown in  figure \ref{fig7}(b). The similarity of the two densities shown in figure \ref{fig7} illustrates the quasi-elastic nature of the collision.

\begin{figure}[!b]

\begin{center}
\includegraphics[width=.6\linewidth,clip]{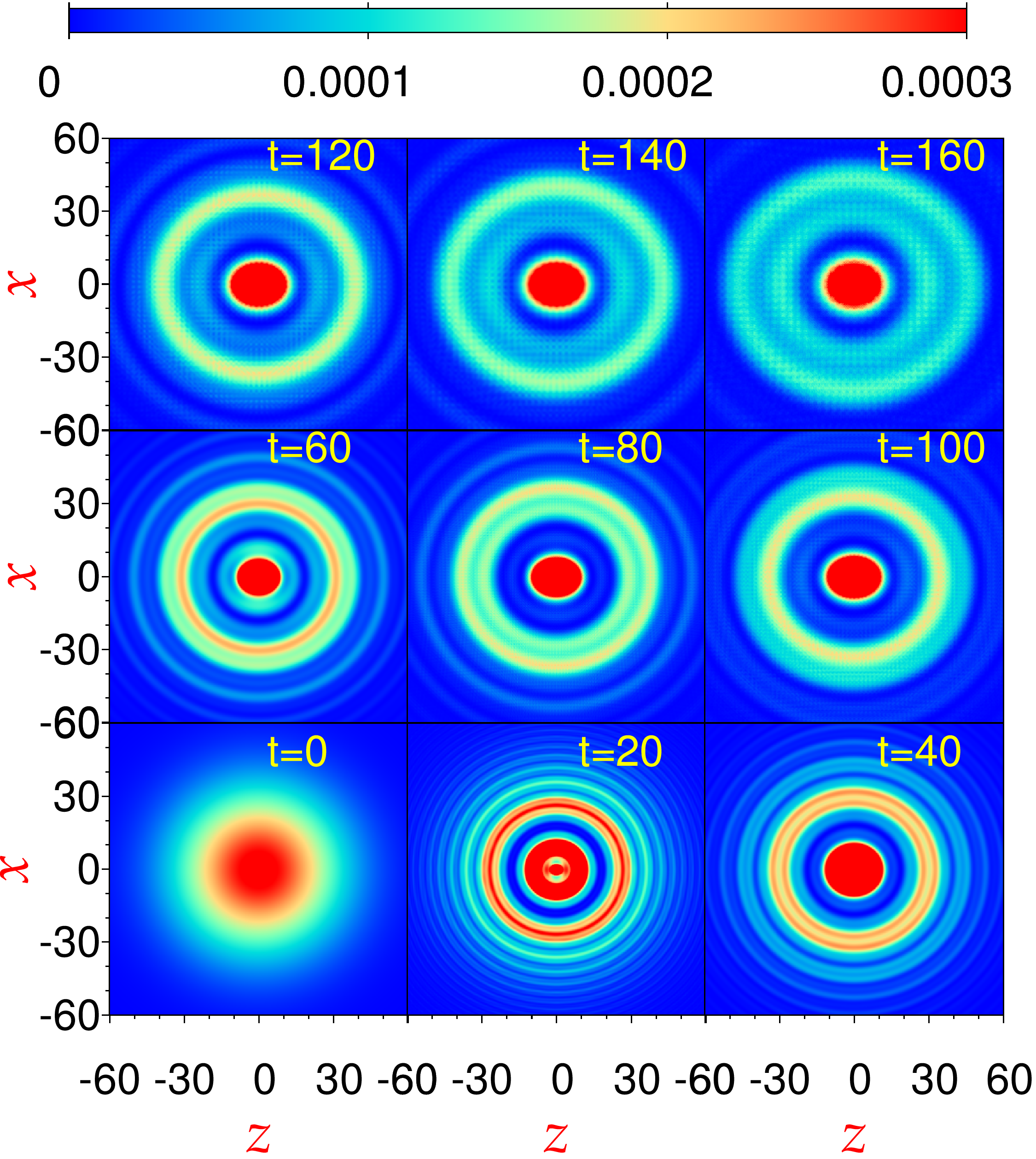}

\caption{ (Color online) The contour plot of density at different times 
during real-time simulation using  Eq. (\ref{eq5}) with $N=1000$ and $a=80a_0$
with 
the  phase imprinted profile 
  (\ref{px}) and (\ref{py}) with $\kappa=0.001$ and $\rho_0=18$
as the initial state.
}\label{fig8} \end{center}

\end{figure}

As the 2D dipolar RDB solitons are stable and robust, they can be prepared from the  phase-imprinted \cite{phase}  Gaussian profile 
\begin{eqnarray}\label{px}
\phi_{2D}(\vec \rho)&=&\sqrt{\frac{\kappa}{\pi}}e^{-\kappa (x^2+z^2)/2}, \quad \rho<\rho_0,\\
 &=& - \phi_{2D}(\vec \rho), \quad \rho\ge \rho_0,\label{py}
\end{eqnarray}
in the trapless case.
In an experiment a homogeneous potential generated by the dipole potential of   a far detuned 
laser beam should be  applied on part of the Gaussian profile  ($\rho<\rho_0$) for an interval of time so as to imprint an extra phase of $\pi$ on the wave function for $\rho<\rho_0$ \cite{dark}. The thus phase-imprinted  Gaussian profile 
with $\kappa =0.001, \rho_0=18$ is propagated in real-time with $N=1000$ and $a=80a_0$, while it slowly transforms into a 2D dipolar RDB soliton. This simulation is done with no trap. In actual experiment a very weak in-plane  trap can be kept during generating 
the 2D RDB soliton, which can be  eventually removed.  The simulation is shown in figure \ref{fig8}, where we show the density at different times. It is illustrated that at large times the  density evolves  towards that of the 2D RDB soliton shown  in figure \ref{fig3}(c).

\section{Summary }

We demonstrated  the possibility of creating a mobile, stable  2D RDB soliton, in  a
 dipolar quasi-2D BEC of $^{164}$Dy atoms polarized along $z$ axis and trapped along $y$ axis,  with a circular notch  and capable of moving in the $x-z$ plane
with  a constant velocity. The 2D RDB solitons are stationary solutions of the mean-field GP equation.
The stability of the RDB soliton is established from a prolonged real-time simulation after
a sudden change of the scattering length.  
The head-on collision between two such solitons with a relative velocity of about 
1.5 mm/s is quasi elastic with the solitons passing through each other with practically no deformation. A possible way of preparing these 2D dipolar RDB   solitons by phase imprinting a Gaussian profile  
is demonstrated using real-time propagation.   The results and conclusions  of this paper can be  tested in experiments with present-day know-how and 
technology  and should lead to interesting future investigations.

We thank the
Funda\c c\~ao de Amparo
\`a Pesquisa do Estado de S\~ao Paulo
(Brazil, Project No. 2012/00451-0) and the Conselho Nacional
de Desenvolvimento Cient\'ifico e Tecnol\'ogico (Brazil, Project  
No. 303280/2014-0) for partial support.

\end{document}